\documentclass[10pt,conference,compsocconf,letterpaper]{IEEEtran}

\ifCLASSINFOpdf
\else
\fi

\usepackage{subfigure}

\usepackage{xspace}
\usepackage{balance}  
\usepackage{graphicx} 
\usepackage{times}    
\usepackage[utf8]{inputenc}
\usepackage[usenames, dvipsnames]{color}
\usepackage{balance}
\usepackage{moreverb}
\usepackage{amsmath}
\usepackage{url}

\usepackage{listings}
\usepackage{color}
\usepackage{textcomp}
\usepackage{algpseudocode}
\usepackage{algorithm}
\usepackage{enumerate}
\usepackage{varwidth}
\usepackage{xcolor}
\usepackage[]{glossaries}

\newcommand{\smartparagraph}[1]{\noindent{\bf #1}\ }

\newcommand{\NprojectName}{$SDI$\xspace}

\usepackage{amssymb}

\definecolor{mygreen0}{rgb}{0, 0.75, 0}

\definecolor{myred1}{rgb}{1,0,0}
\definecolor{mygreen1}{rgb}{0, 1, 0}
\definecolor{myblue0}{rgb}{0, 0, 1}

\definecolor{myred2}{rgb}{1,0.5,0.5}
\definecolor{mygreen2}{rgb}{0.5, 1, 0.5}
\definecolor{myblue2}{rgb}{0.5, 0.5, 1}

\definecolor{mygreen}{rgb}{0, 0.25, 0}
\definecolor{myblue}{rgb}{0, 0, 0.75}
\definecolor{myred0}{rgb}{0.5,0,0}

\definecolor{listinggray}{gray}{0.98}
\definecolor{lbcolor}{rgb}{0.98,0.98,0.98}
\begin{document}
%
\bibliographystyle{IEEEtran}
\title{Latency-Sensitive Web Service Workflows: \\A Case for a Software-Defined Internet}


\author{\IEEEauthorblockN{Pradeeban Kathiravelu}
\IEEEauthorblockA{Emory Univeristy\\
Atlanta, GA, USA\\
Tel: (+1) 404 784 9117\\
pradeeban.kathiravelu@emory.edu}
\and
\IEEEauthorblockN{Peter Van Roy}
\IEEEauthorblockA{Université catholique de Louvain\\
Louvain-la-Neuve, Belgium\\
peter.vanroy@uclouvain.be}
\and
\IEEEauthorblockN{Lu{\'\i}s Veiga}
\IEEEauthorblockA{Instituto Superior Técnico\\
Universidade de Lisboa\\
Lisboa, Portugal\\
luis.veiga@inesc-id.pt}
\and
\IEEEauthorblockN{Elhadj Benkhelifa}
\IEEEauthorblockA{Staffordshire University\\
Staffordshire, UK\\
e.benkhelifa@staffs.ac.uk}
}

\maketitle

\begin{abstract}

The Internet, at large, remains under the control of service providers and autonomous systems. The Internet of Things (IoT) and edge computing provide an increasing demand and potential for more user control for their web service workflows. Network Softwarization revolutionizes the network landscape in various stages, from building, incrementally deploying, and maintaining the environment. Software-Defined Networking (SDN) and Network Functions Virtualization (NFV) are two core tenets of network softwarization. SDN offers a logically centralized control plane by abstracting away the control of the network devices in the data plane. NFV virtualizes dedicated hardware middleboxes and deploys them on top of servers and data centers as network functions. Thus, network softwarization enables efficient management of the system by enhancing its control and improving the reusability of the network services. In this work, we propose our vision for a Software-Defined Internet (\NprojectName) for latency-sensitive web service workflows. SDI extends network softwarization to the Internet-scale, to enable a latency-aware user workflow execution on the Internet.

\end{abstract}


%
\IEEEpeerreviewmaketitle



%
\newglossarystyle{my}{%
  \setglossarystyle{long3colheader}%
  \renewcommand*{\glossaryheader}{%
    \toprule
    Acronym & Description & Page \tabularnewline\midrule\endhead
    \bottomrule\endfoot
  }
}

\setglossarystyle{my}
\renewcommand*{\glossaryname}{List of Acronyms}
\renewcommand*{\entryname}{Acronym}
\renewcommand*{\descriptionname}{Description}
\renewcommand*{\pagelistname}{Page}

\renewcommand*{\glsnamefont}[1]{\bfseries #1}%
 
\newcommand{\subscript}[2]{$#1 _ #2$}

\section{Introduction}
\label{sec:intro}

While the Internet continues to grow organically, it relies on Internet hubs to connect the rest of the world. These Internet hubs are geographically far from the end users belonging to the rest of the world, scattered across the globe. Consequently, a large share of Internet users suffers from high latency and jitter. On the other hand, Internet services are getting diverse and pervasive, with several third-party networks, cloud, and service providers offering resources to the end users. Edge environments continue to replace cloud platforms for latency-sensitive web applications~\cite{shi2016edge}, such as eScience workflows and the IoT~\cite{de2018application, zhang2015cloud, villari2017software}, due to their increasing demand for a quick response. A complex user workflow often requires consuming several resources from numerous providers~\cite{abrishami2013deadline}. Despite the growing number of service providers, users still cannot seamlessly choose services from multiple providers to compose their workflows, due to the incompatibility between the service providers. Increasing volume and variety of the providers, yet with lack of interoperability among their interfaces~\cite{dillon2010cloud}, makes composing service workflows abiding by the user policies a hard problem.

Through its unified view and control of the data plane devices, SDN~\cite{mckeown2009software} facilitates programmability and management capabilities to the network. SDN separates the control of the data plane devices into a logically unified network controller. SDN facilitates a global awareness of the network data plane devices, typically within a cloud or a data center, but also extended to Wide Area Network (WAN) scenarios such as Content Delivery Networks (CDNs)~\cite{wichtlhuber2015sdn}. On the other hand, NFV~\cite{han2015network} virtualizes various network services and deploys them on servers as Virtual Network Functions (VNFs)~\cite{bari2015orchestrating} instead of having them as individual hardware middleboxes~\cite{walfish2004middleboxes}. VNFs as software middleboxes are cheaper to acquire and easier to manage from a global controller, compared to hardware middleboxes. These traits have indeed facilitated the adoption of network softwarization by several service and network providers~\cite{feamster2013road}.

The demand for a locality-aware execution, where the infrastructure and the service executions move closer to the user~\cite{bonomi2012fog}, is met with an increasing number of edge providers, to serve the large and geographically-distributed user base~\cite{villari2016osmotic}. Cloud providers themselves are opening up more regions~\cite{zhang2010cloud}, to offer better Quality of Service (QoS) to the remote users. However, lack of interoperability across the providers hinders a wide-scale adoption of the prevalent distributed cloud and edge environments, as users still cannot select services from multiple providers to compose their workflows.

The Internet connects the entire globe as a single large-scale network. As a network, the Internet is built by the interconnection of multiple networking systems of all the nations. The organizational policies and the bilateral relationships of the countries enable peering contracts between the autonomous systems of the Internet. For example, two nations without a bilateral relationship often cannot interconnect directly. They have to go through a third-country to communicate despite their geographical proximity. We posit that exploiting the edge resources at the application layer to create an overlay to complement the Internet, rather than using the default Internet paths, may provide a low-latency alternative.

We need to complement classic SDN with more widespread light-weight Service Oriented Architecture (SOA) such as Message-Oriented Middleware (MOM)~\cite{curry2004message} protocols, to achieve interoperability and coordination across service executions at Internet scale. Existing approaches that exploit application layer protocols to manage the networks together with SDN~\cite{phemius2014disco} are still limited in scalability and interoperability, with little support for diverse application scenarios such as IoT and big data workflows. SOA offers scalable distributed executions across wide-area multi-domain networks, supporting a vast range of devices and services. We should extend SDN with SOA to facilitate efficient management of heterogeneous services, rather than just the network infrastructure.

This paper presents Software-Defined Internet (\NprojectName), an approach that extends network softwarization to the Internet, aiming to provide a latency-aware web service workflow execution leveraging the Internet latency measurements from the user devices. \NprojectName extends network softwarization to solve the challenges of scheduling web service workflows at an Internet-scale. The current workflow placement approaches are limited in terms of feasibility, scalability, and optimality in efficiently provisioning resources for user workflows spanning various infrastructure and service providers across the Internet. The diversity of services and their users make an optimal service composition and workflow placement, for a tenant user consuming several third-party services, a complex research problem~\cite{klein2012towards}. We first identify several challenges that should be addressed to be able to build \NprojectName to schedule latency-sensitive web service workflows. We posit such web service workflows as a case for an \NprojectName approach that extends and leverages network softwarization. \NprojectName measures Internet bandwidth across the countries in real-time and adopts the web service workflow scheduling accordingly.

In the upcoming sections, we present our vision for an \NprojectName and discuss our approach. We discuss our motivation for an SDI in Section~\ref{sec:motivation}. We then elaborate the \NprojectName approach in Section~\ref{sec:approach}. We further present our findings in Section~\ref{sec:eval}, and discuss the related work in Section~\ref{sec:related_work}. Section~\ref{sec:conclusion} finally concludes the paper with a summary and findings.

\section{Background and Motivation}
\label{sec:motivation}

In this section, we look into the rationale behind the asymmetry of the Internet cost and performance, and how the users may be able to mitigate the high latency for their service workflow executions using the Internet measurements and network softwarization.

Several factors influence the performance and cost of interconnections, locally and across countries. Rural areas have an asymmetric cost due to the low user density compared to the cost associated with extending the network beyond the urban areas. Such additional costs are subsidized, paid by the provider often with minimal income, or paid by the end user, thus leading to a much higher bandwidth cost compared to their urban neighbors ~\cite{dymond2004telecommunications}. On a grand scale, such asymmetry expands to incur higher bandwidth costs for developing countries compared to the major Internet hubs. For instance, the northern hemisphere tends to have more established interconnections, and consequently, cheaper and faster than the southern hemisphere~\cite{jensen2005interconnection}.

Network softwarization has yielded positive outcomes concerning the performance and management of the network architectures while minimizing the capital and operational expenditures (CapEx and OpEx) of the enterprises~\cite{bouras2016cost}. Two major technological factors drive the preference for network softwarization: i) performance efficiency and flexibility achieved by separating the network infrastructure from the execution~\cite{hwang2015netvm}, and ii) the ability to control the network flows based on the user preferences from the application plane, for a high QoS~\cite{taleb2017permit}. SDN controllers are, in practice, software applications developed in high-level languages, such as Java or Python. Therefore they can be extended and invoked from the application layer. They, on the other hand, control and manage the network data plane devices. Thus, the controllers are capable of providing cross-layer optimizations to the network systems, by receiving status updates from the network plane, while adhering to the policies specified from the application plane. While SDN can offer network-awareness through the unified view of its controller, it typically limits its scope to a data center. Consequently, several challenges in separating a large-scale network environment consisting of multiple domains from its infrastructure remain unaddressed. While such multi-domain workflows are still uncommon, we observe that the growing adoption of 5G and IoT will quickly bring the need for such workflows.

\begin{figure}[!ht]
    \begin{center}
            \vspace{-0.5em}

        \resizebox{\columnwidth}{!}{
            \includegraphics[width=\textwidth]{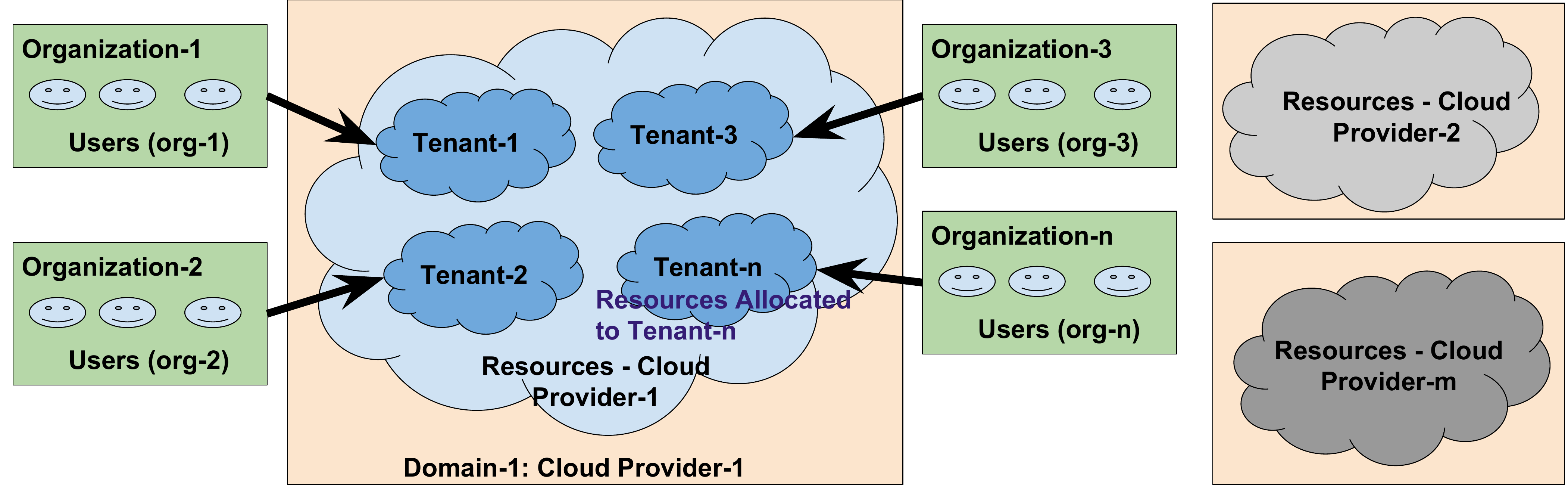}
        }
    \end{center}
    \caption{Multitenancy and the Tenant Users of a Cloud Environment}
            \vspace{-0.5em}

    \label{fig:thesismultitenancy}
\end{figure}

The cloud environments consist of users from several organizations. We call these sets of users the \textit{tenants} of the systems. Multitenancy~\cite{fiaidhi2012enforcing}, the ability to support several tenants with shared resources, is a core pillar of cloud computing. It advocates sharing of the underlying infrastructure and platform among several third-party organizations, i.e., the tenants of the environment. A tenant typically consists of a set of users controlled by a single administrator account of the organization. Each tenant receives its own `slice' of the cloud resources without sacrificing the privacy and isolation of data and execution belonging to each tenant, as illustrated by Figure~\ref{fig:thesismultitenancy}. However, multitenancy comes with the cost of limited control to the end users -- the resources are entirely managed and provisioned by the provider, often oblivious of the sophisticated end user policies. Furthermore, each cloud environment is maintained by its provider as a network domain, independent and often incompatible with other cloud environments. Therefore, the users have limited control in resource allocation in multi-domain environments (consisting of several cloud and edge providers), considering all the available providers.

The Internet is an extensive network of networks that connects the data centers, servers, tenants, and users across the globe~\cite{bailey1997economics}. Naturally, the Internet is also the largest network in the world. Internet bandwidth and latency are crucial aspects in deciding how the residents of a country communicate with the outside world. The Internet bandwidth, or the data rate, between regions is not only determined by the physical cables between the regions. It is also heavily influenced by the interconnection agreements~\cite{drpeering} across the connectivity providers such as the Internet Service Providers (ISPs) and transit providers~\cite{ma2010cooperative}. Predominantly, computer research on the Internet-scale has focused on technological aspects. The existing research does not adequately study the Internet connectivity and geopolitical factors.

Human interactions and policy decisions play a significant role in deciding the interconnections, whether to establish direct connectivity between two autonomous systems (ASes) of the Internet~\cite{gao2001inferring}. The Internet is built not only on computer networks and servers but also on interactions between organizations. The interconnection agreements between two ASes can be either transit or peering. In a transit agreement, an ISP typically pays to a transit provider, to reach the rest of the Internet. On the other hand, peering agreements are made between two relatively equal ASes to share their downstream customers. Peering agreements commonly do not involve a monetary transaction, primarily if neither of the peering ASes asymmetrically relies on the other AS for most of its interconnection.

Strategically developing such interconnection agreements is crucial for the growth of the AS. The relationships between the ASes are a key factor in deciding the Internet interconnections. Despite being physically close, two countries or regions may share poor Internet connectivity due to various reasons. The primary factors include lack of interconnection agreements between the autonomous systems from these regions driven by lack of demand and economic reasons~\cite{bailey1997economics}, political factors that prevent direct communications, and lack of infrastructure to facilitate such interconnection. Developing local interconnection agreements are essential for faster Internet access in an efficient manner, for the regions as well as for the Internet itself globally.

The Internet follows a hub-and-spoke topology with several major Internet hubs~\cite{grubesic2003geographic}. Most Internet traffic is routed through these hubs. This approach is similar to the air traffic, where each career has its own hub, with most international flights routed through those airports. While such an approach helps with air travel at several levels, the move to the hub-and-spoke topology on the Internet is entirely driven by economic reasons. The ASes focus more on the hubs, thus avoiding giving equal importance to geographical locations with less demand or user density (i.e., spokes). However, due to this topology, remote Internet regions suffer a high latency even to access content from nearby regions. Furthermore, such a topology increases the long-haul Internet traffic, making remote Internet regions unfit for latency-sensitive web applications. Increasing interconnections between neighbor regions can minimize such dependencies, thus reducing latency.

The current services ecosystem gives limited flexibility to the tenants who share the network for their services placement and execution. This limited flexibility and control given to the tenants prevent them from leveraging multiple service providers at the cloud and the edge for a single workflow, even when the tenants possess the ability to develop and deploy network applications seamlessly across various execution environments. Even the current network softwarization approaches such as classic SDN give a limited capability for the tenant applications to pass their preferences and policies to the network level. This state of affairs limits the tenants from efficiently ensuring Service Level Objectives (SLOs) at the network level, more so when the network is managed by a third-party and shared by several other tenants.

The economic and geopolitical aspects of the Internet requires additional study. The Internet is an evolving entity. On-going major political events may have an impact on the Internet connectivity of the region. For example, Internet censorships enforced by governments. The Internet connectivity metrics published by the service providers are static, and they are unable to cope with such dynamic changes. On the other hand, distributed networks deployed on the Internet can be used to measure Internet performance in real-time. RIPE Atlas offers access to several computer servers deployed in almost all the countries~\cite{bajpai2015lessons}. Using RIPE Atlas clients, we can identify the Internet latency between two Internet endpoints, known as RIPE Atlas probes. We look into the interconnection performance between pairs of countries and regions on the Internet as a motivation for an \NprojectName approach. Furthermore, while the Internet paths do not consider latency, with latency information readily available, we posit \NprojectName as a solution to deploy web service workflows that leverage Internet paths as well as direct connections across the cloud and data center nodes. \NprojectName extends network softwarization to the Internet, through an application layer network overlay. It thus provides better flexibility and control for the tenant workflows and offers low latency to the user web service workflows on the Internet.

\section{Solution Architecture}
\label{sec:approach}

Our proposed unified \NprojectName framework aims to bring the control of scheduling the web service workflows to the user by enhancing the application layer to route the workflows in a latency-aware manner. \NprojectName uses our previous work, Évora~\cite{kathiravelu2018composing}, to find the optimal set of services to compose the user workflows. Évora exploits network softwarization to construct workflows at the edge, adhering to the user-defined workflow scheduling policies.

\subsection{\NprojectName Approach}
As the core enabler and prerequisite of our framework, we first must ensure interoperability across the diverse execution environments as well as the variety of tenant workloads, for seamless scheduling and migration of service workflows. The services composing the workflows should be interoperable, to execute service workflows spanning multi-domain networks managed by multiple providers, with minimal repetitive development and manual deployment efforts from the network application developers.

The current cloud ecosystem consists of mostly disconnected networks with little interaction and coordination across the infrastructure and service providers at the cloud and the edge. The ability to compose a workflow spanning multiple providers is limited by not only the technical challenges but also the business vision and the enterprise policies of the providers. \NprojectName separates the network from the infrastructure, to counter the dependence of the network service workflows on the providers. It facilitates an independent third-party user to consume resources from multiple cloud or edge providers seamlessly, using an overlay network on top of the connected cloud and edge networks. This inter-cloud architecture enhances the interoperability of the cloud infrastructure further, while also providing performance enhancements to the tenant cloud applications.

\NprojectName enriches the services ecosystem with its core contributions while having the aforementioned interoperability enhancements in place as a prerequisite and foundation for network-aware web service workflow scheduling across the multi-domain networks. By using messages for inter-domain coordination, rather than static dedicated network links, the extended SDN architecture ensures not to introduce hierarchies among the networks, or compromise their security. Using a MOM approach facilitates multi-domain control via dedicated network connections as well as the public Internet with protected access, thus further aiming to enhance the performance, scalability, and extensibility of the multi-domain service workflow compositions and migrations. \NprojectName thus enables composing workflows on services deployed on multiple network domains and migrates the execution based on the load on the service instances.

While our enhancements enable composing network service chains, consuming the services deployed on multiple edge and cloud providers, we must compose and schedule these web service workflows adhering to the user policies and SLOs. Remarkably, workflows have additional constraints compared to the stand-alone service instance selection and execution. We incorporate Évora service composition and workflow scheduling algorithms into \NprojectName to ensure that the service instances are chosen while adhering to both user policies and resource availabilities at the nodes. \NprojectName uses both the network-level statistics facilitated by the SDN architecture extended with MOM, as well as the service-level statistics from the web services engines.

\subsection{\NprojectName Prototype}

\begin{figure}[ht]
        \vspace{-1em}

    \begin{center}
        \resizebox{0.8\columnwidth}{!}{
            \includegraphics[width=0.8\textwidth]{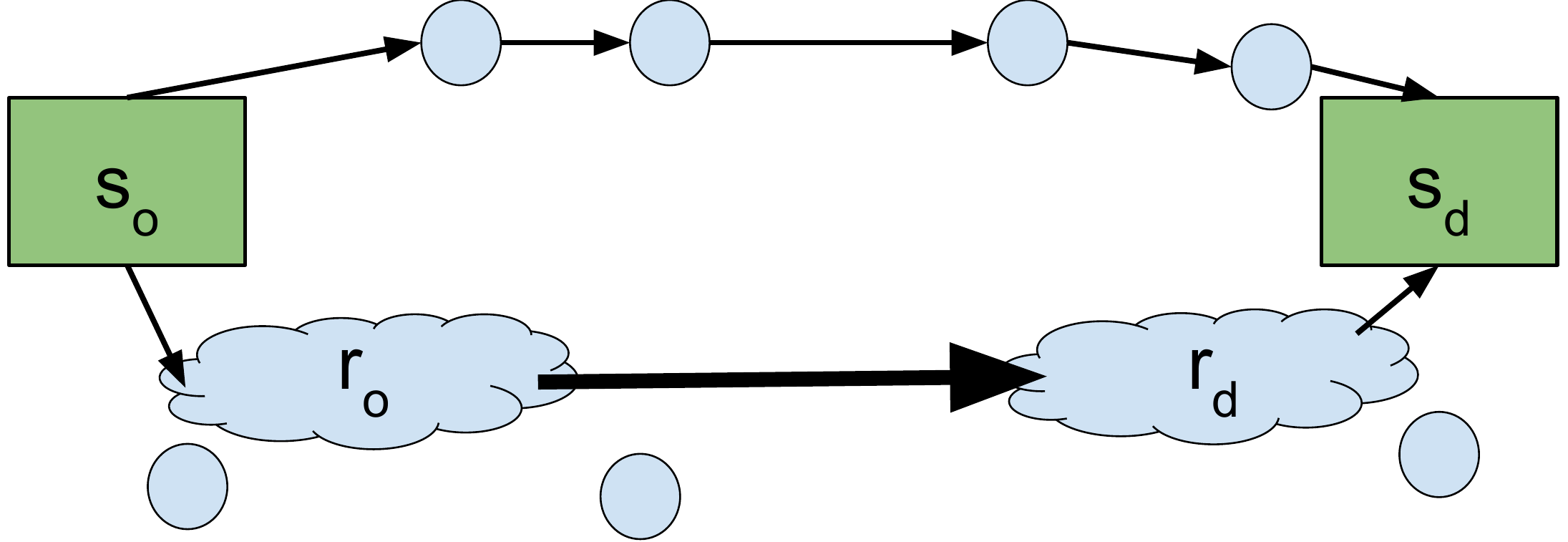}
        }
    \end{center}
    \vspace{-1.5em}
    \caption{\NprojectName Deployment}
        \vspace{-0.5em}

    \label{fig:deplarch}
    
\end{figure}

Figure~\ref{fig:deplarch} represents a sample \NprojectName deployment that sends data between two endpoints $s_o$ and $s_d$. $s_o$ is the origin server, where the data transfer originates. It is close to the cloud region $r_o$. It is connected to the Internet via an ISP. $s_d$ is the destination server, which is close to the cloud region $r_d$. A cloud overlay connects the regions $r_o$ and $r_d$. We connected $s_o$ to $r_o$ via the public Internet (thus with a throughput of $T(s_o,r_o)_I$). Throughput of \NprojectName is limited by $max(T(s_o,r_o)_I, T(r_o,r_d)_C, T(r_d,s_d)_I)$. Here, $T(s_o, r_o)_I)$ and $T(s_d, r_d)_I)$ refer to the data rate of connecting $s_o$ and $s_d$ to their respective nearby cloud regions $r_o$ and $r_d$ via the public Internet. $T(r_o, r_d)_C)$ refers to the data rate of \NprojectName overlay between the cloud regions, as offered by the cloud provider.

Due to its reliability and access, RIPE Atlas~\cite{bajpai2015lessons} has become a prominent resource for Internet measurements research. Each \NprojectName origin is occupied with credits from RIPE Atlas to make Internet measurements. Attaching a RIPE Atlas to the router of the endpoint server will provide more accurate latency measurements. However, when a probe is not present at the endpoint, using the average from the other probes in the region offers a sufficient estimate. 

By hosting a RIPE Atlas probe ourselves, we accumulated more than 10 million RIPE Atlas credits. We spent the credits on measuring the Internet performance on an enormous scale, continuously between selected endpoints. We posit that by leveraging network softwarization with a focus on the end user devices, the end users can gain more control over the network and their applications in a multi-tenant environment. However, supporting user policies through the separation of network flows, and creating virtual execution spaces efficiently, remain open challenges to be addressed. We observe that to fully reap the benefits of the pervasive edge nodes and network softwarization, we should bring the control of the resource allocation and executions back to the users, despite sharing the resources from multiple providers across several tenants.

\begin{figure}[ht]
    \begin{center}
            \vspace{-1em}

        \resizebox{0.65\columnwidth}{!}{
            \includegraphics[width=0.65\textwidth]{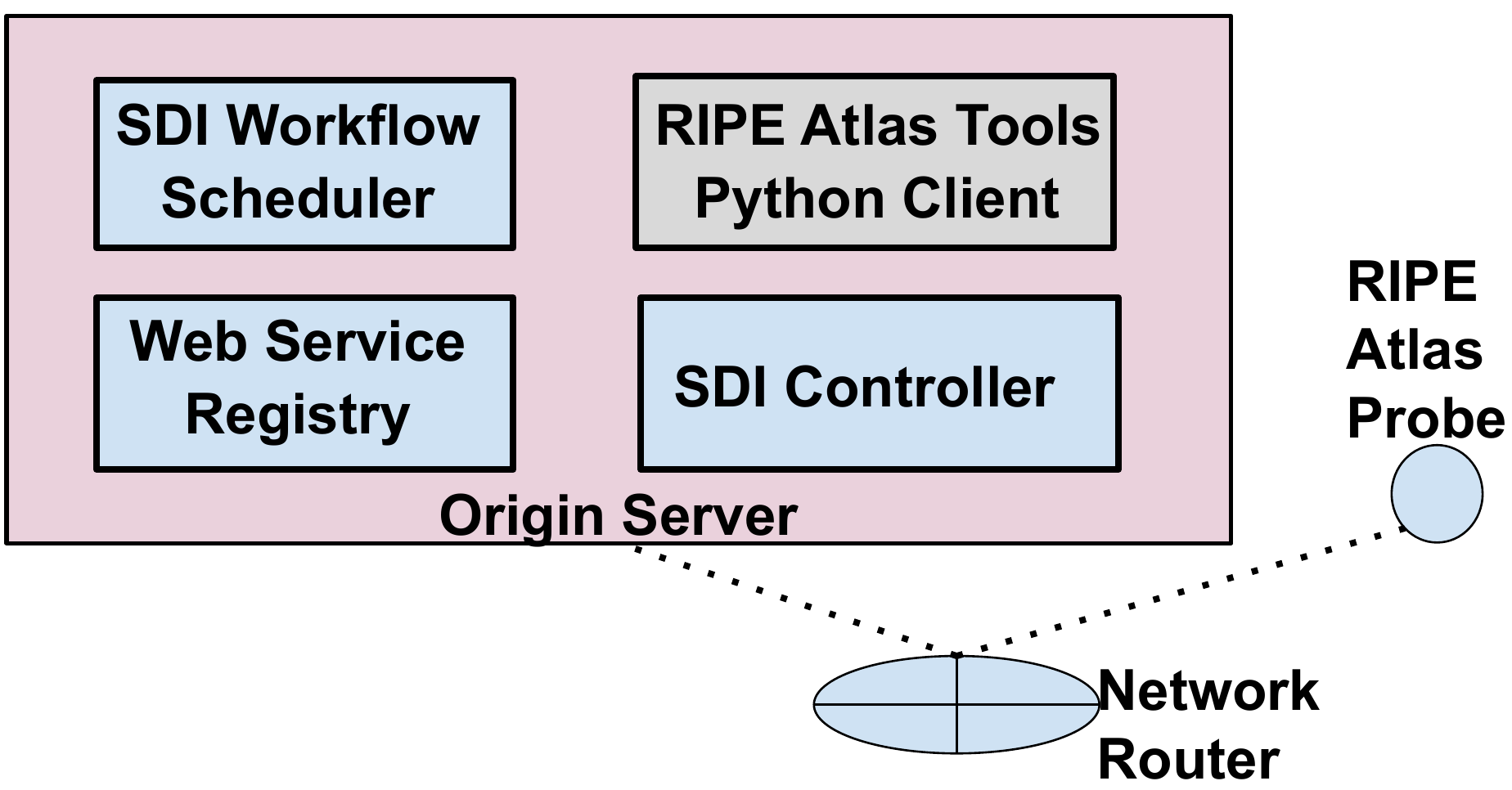}
        }
    \end{center}
                \vspace{-1em}

    \caption{\NprojectName Solution Architecture}
                \vspace{-0.5em}

    \label{fig:solarch}
\end{figure}

Figure~\ref{fig:solarch} illustrates the solution architecture of \NprojectName. RIPE Atlas consists of a command-line tool~\cite{ripetools} that is integrated with the application layer of the web service client. Therefore, the Internet measurements are readily available for the \NprojectName workflow scheduler, powered by Évora. The origin server consists of the \NprojectName controller, which manages the data transfer via the Internet or the alternative cloud and edge paths. We use the Python client tools provided by the RIPE Atlas to find the Internet latency across various nodes and regions. The \NprojectName Workflow Scheduler then finds the nodes to send the data, based on the Web Service Registry. The Web Service Registry consists of the list of web service nodes. Each deployment optionally consists of a RIPE Atlas Probe connected to the same router as the origin server, to provide accurate latency information to the other participating nodes and to accumulate credits to perform Internet measurements.

\begin{figure}[ht]
    \begin{center}
        \resizebox{\columnwidth}{!}{
                        \vspace{-2em}

            \includegraphics[width=\textwidth]{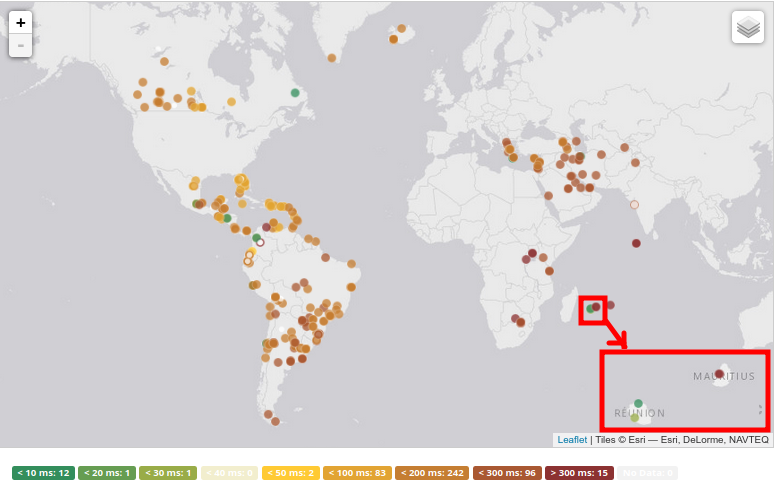}
        }
    \end{center}
                \vspace{-1em}

    \caption{RTT Measurements with RIPE Atlas}
                \vspace{-0.5em}

    \label{fig:rtt}
\end{figure}

To highlight the case for latency variations across geographical regions, we measured the round-trip time (RTT) and the number of hops with RIPE Atlas traceroute measurements. We sent data packets every hour, for 33 days, using 6.8 million RIPE Atlas credits in total~\cite{mymeasure}. Our destination node is a RIPE Atlas probe in Colombia. We successfully used 491 RIPE Atlas probes scattered across the globe as the origin nodes. We continuously sent data from the origin probes to the destination probe. Figure~\ref{fig:rtt} depicts the RTT of the data transfers. We observe that the RTT does not necessarily depend on the geographic distance, as can be seen by a lower RTT for a few pairs of probes that are geographically far from each other, compared to a few geographically closer pairs. For example, low RTT was observed when data was sent between Réunion and Colombia. However, despite its geographical proximity to Réunion, Mauritius displayed a much larger RTT to Colombia. In the presence of direct connections and cloud overlays between Réunion and Mauritius, such slow performance can be rectified with the \NprojectName approach.

\begin{figure}[ht]
    \begin{center}
        \resizebox{\columnwidth}{!}{
            \includegraphics[width=\textwidth]{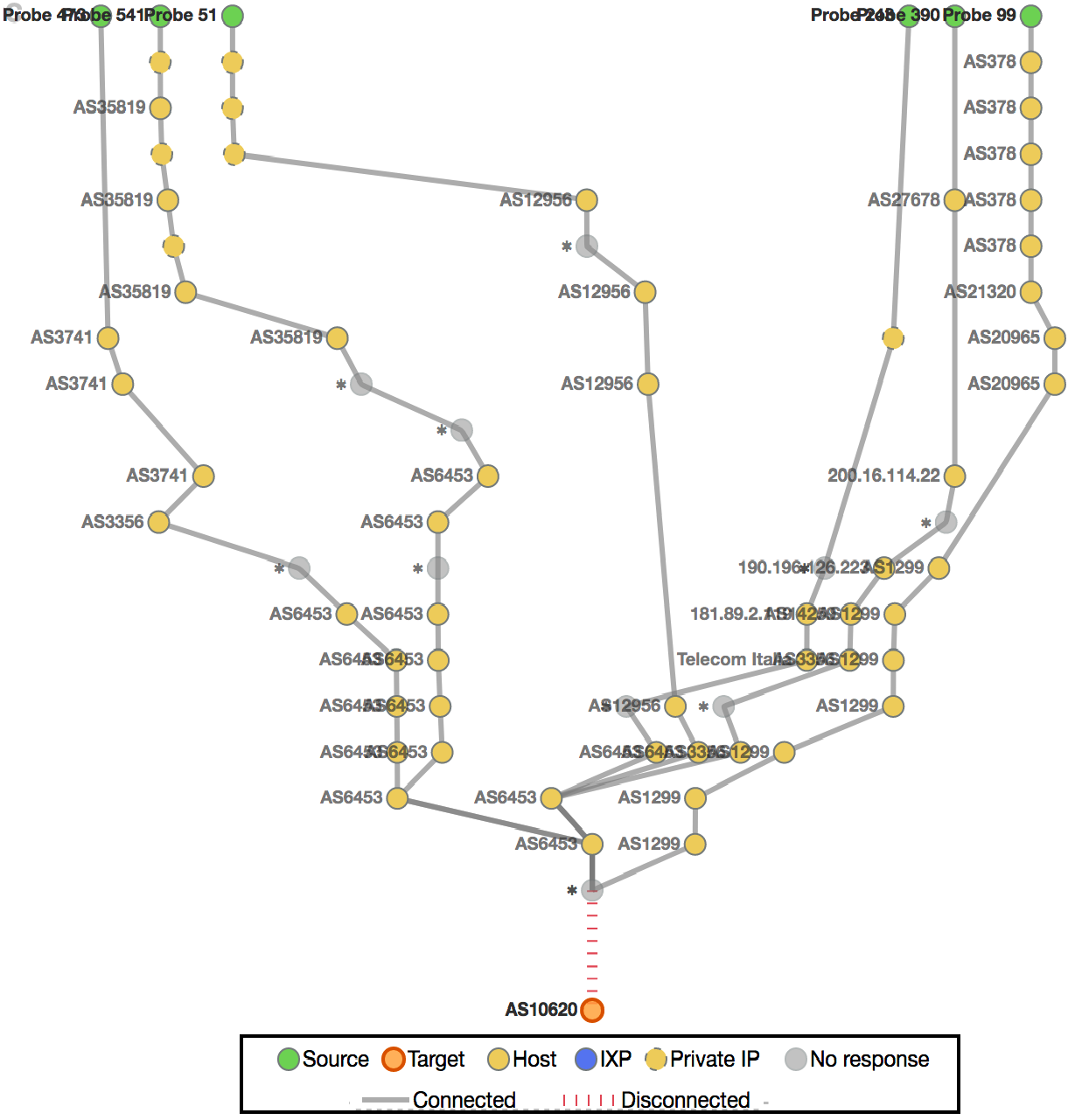}
        }
    \end{center}
    \vspace{-1em}
    \caption{Traceroute visualization with RIPE Atlas}
    \vspace{-2em}
    \label{fig:traceroute}
\end{figure}

Figure~\ref{fig:traceroute} visualizes the traceroute of the data transfer to the destination probe from 6 of the 491 origin probes. RIPE provides this visualization and the corresponding data in real-time. While this information is specific to the AS, \NprojectName estimates the performance of a region as an average from all the probes/ASes belonging to the region. While we built our \NprojectName prototype based on RIPE Atlas measurements, an \NprojectName implementation can use any latency-aware overlay network. 

\section{Preliminary Evaluation}
\label{sec:eval}
We modeled a prototype of \NprojectName that utilizes direct interconnections between two Internet endpoints in a latency-aware manner. Due to their availability, we used AWS cloud instances to replace a large segment of the Internet path between two endpoints. We used RIPE Atlas probes as the endpoints. By configuring the network path at the application layer with RIPE Atlas measurements, we ensured the data transfer across the endpoints, transferring through the cloud endpoints. We benchmarked latency, latency variations as jitter, throughput, and loss rate of \NprojectName against that of the public Internet entirely based on ISP-based connectivity.

\begin{figure}[ht]
    \begin{center}
        \resizebox{0.9\columnwidth}{!}{
            \includegraphics[width=0.9\textwidth]{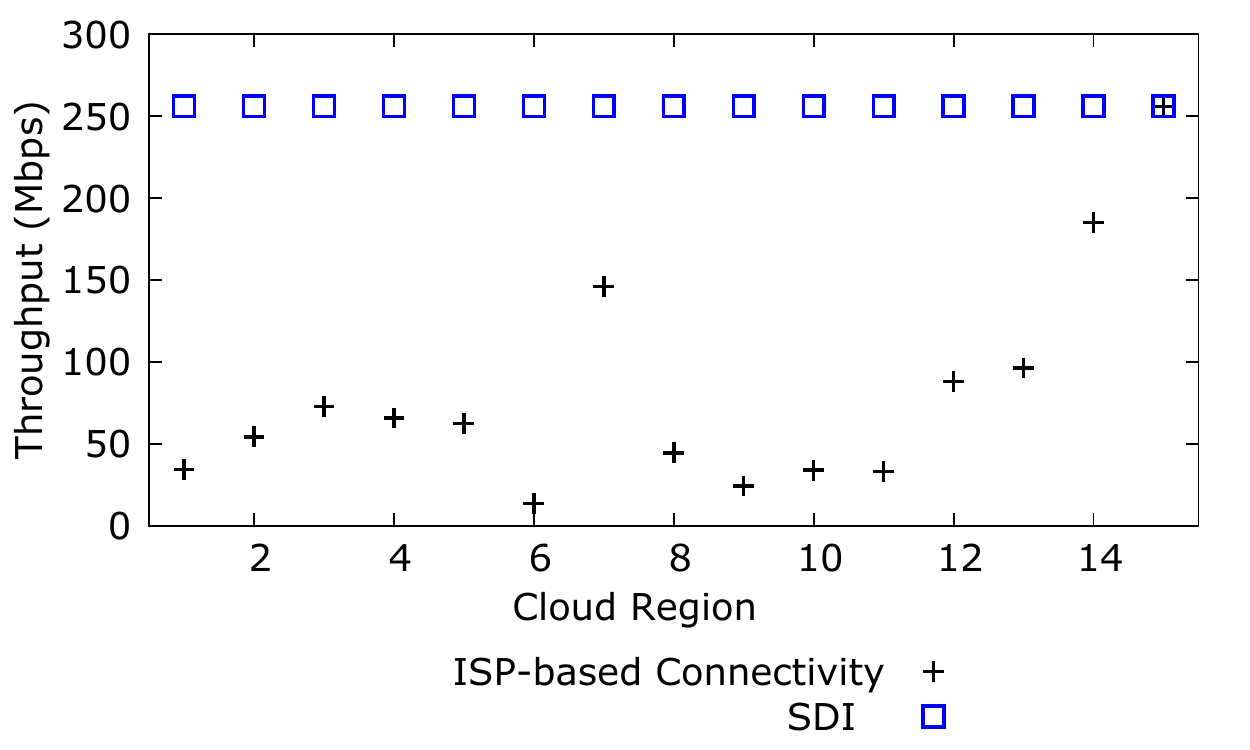}
        }
    \end{center}
            \vspace{-1.5em}

    \caption{Throughput of \NprojectName with ISP-based cloud connect}
            \vspace{-1.5em}

    \label{fig:netuberthroughput}
\end{figure}

First, we measured the throughput of \NprojectName against that of using the public Internet access provided by the ISPs, in transferring data between two distant endpoint servers, $s_o$ and $s_d$. We measured the throughput of data flows between the endpoints directly via the public Internet. Then we measured throughput of \NprojectName, via \NprojectName cloud overlay between $r_o$ and $r_d$. Figure~\ref{fig:netuberthroughput} illustrates the achieved throughput in sending data from our server in Atlanta to cloud servers in various regions, first directly via the ISP-based connectivity, and then via \NprojectName. \NprojectName sends the data first to the region that offers the highest data rate between our server and the cloud. Then, the data is forwarded via the \NprojectName cloud overlay to the cloud server in the destination region. We observed North Virginia offer the highest data rate in our experiment, as it provided 256 Mbps when connected to our server via the ISP. We note the geographical proximity of our server to the North Virginia region as a potential influencing factor in providing high throughput. The cloud overlay offered 1.2 Gbps. Therefore, the path between the origin server and the origin cloud region remains the bottleneck in \NprojectName data transfer.

We observe that connecting the cloud servers from far regions via the ISP offered lower throughput. ISP and \NprojectName provide the same throughput when $r_o$ = $r_d$, i.e., when the destination server is in the closest region to the origin server (as can be seen by the cloud region 15 in Figure~\ref{fig:netuberthroughput} which refers to North Virginia). By routing the data traffic via the high-performing path consisting of Atlanta $\rightarrow$ North Virginia, we exploit \NprojectName to offer a higher end-to-end data rate compared to ISP. \NprojectName utilizes the cloud path as well as the faster connectivity to the nearest cloud region to provide a uniform data rate to the cloud servers, regardless of their regions.

\begin{figure*}[!ht]
        \centering
        \subfigure[Atlanta to Sydney\label{fig:jitter-a}]{\includegraphics[width=0.45\textwidth]{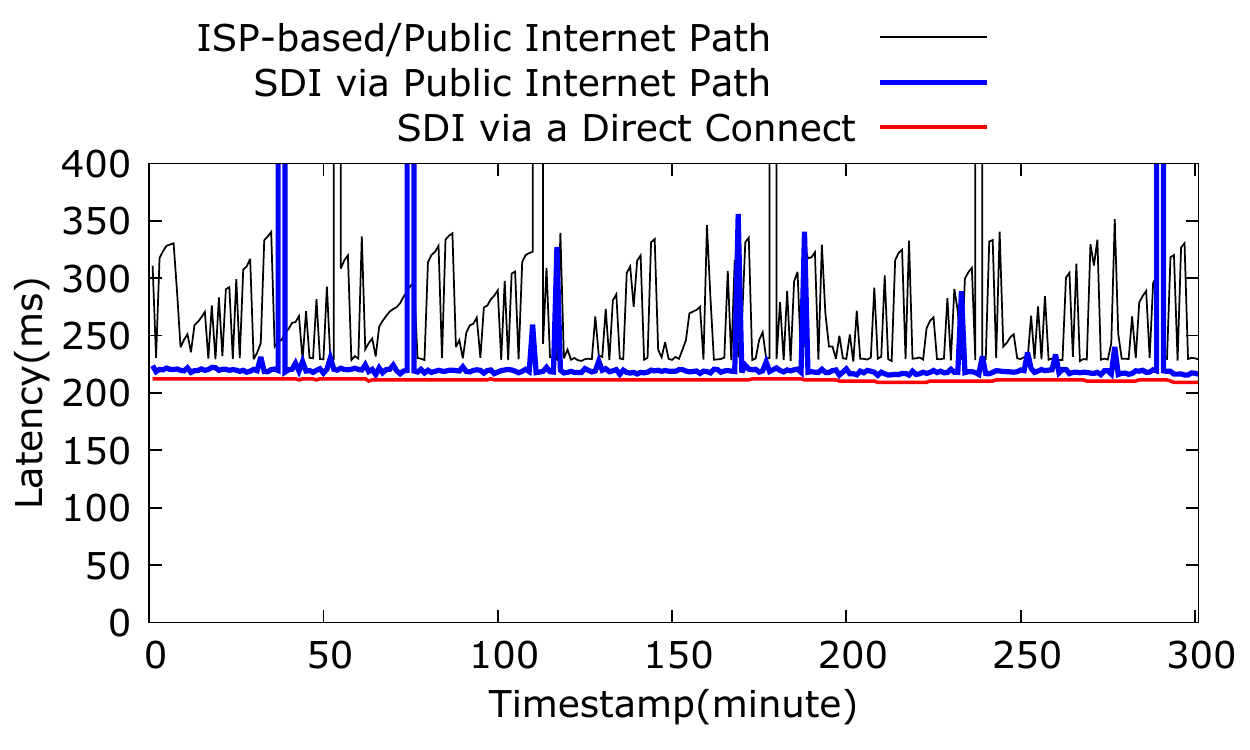}                              \vspace{-1em}
}
          \subfigure[Atlanta to Tokyo\label{fig:jitter-b}]{\includegraphics[width=0.45\textwidth]{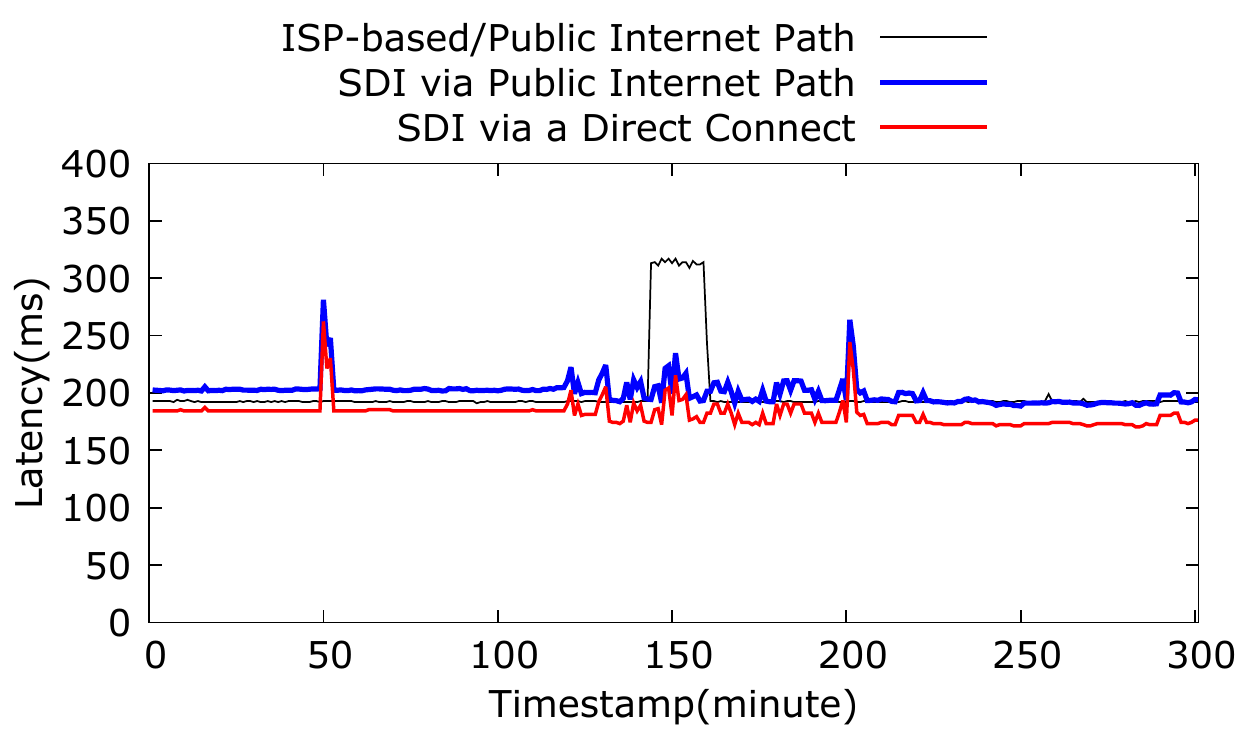}                              \vspace{-1em}
}\\                   
                              \vspace{-4em}

          \subfigure[Atlanta to Mumbai\label{fig:jitter-c}]{\includegraphics[width=0.45\textwidth]{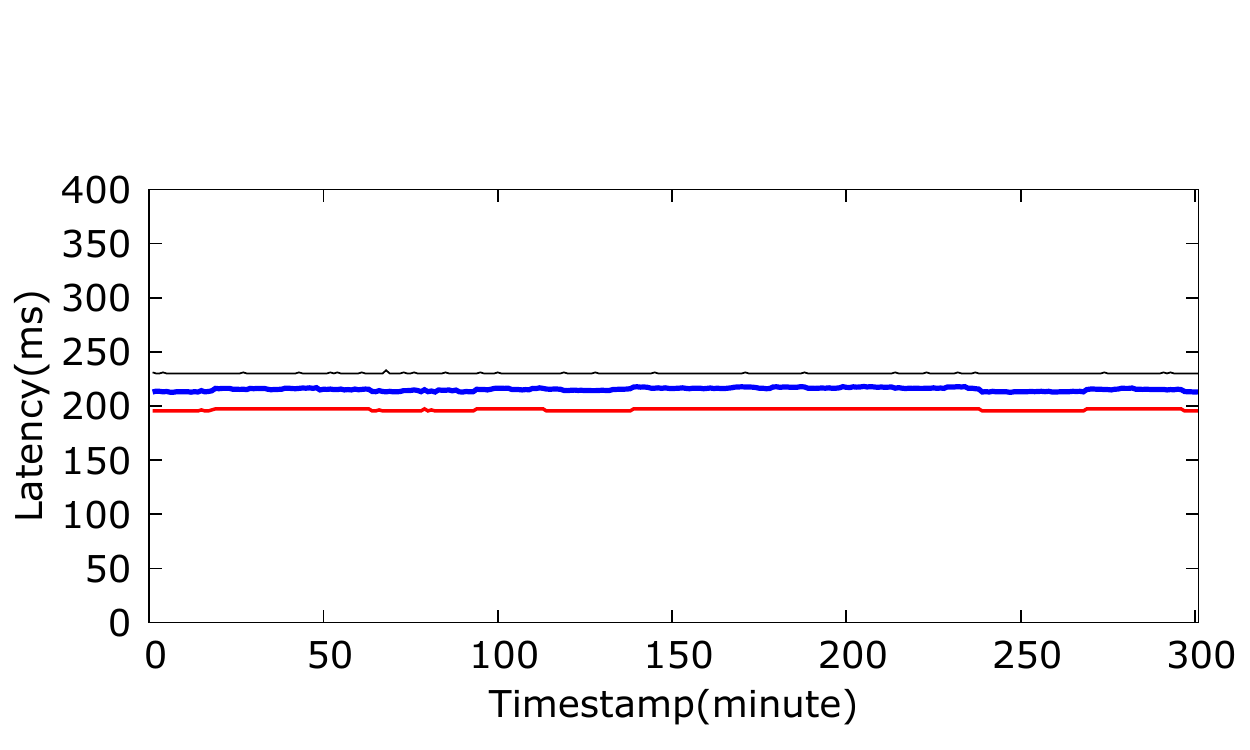}}
          \subfigure[Atlanta to Seoul\label{fig:jitter-d}]{\includegraphics[width=0.45\textwidth]{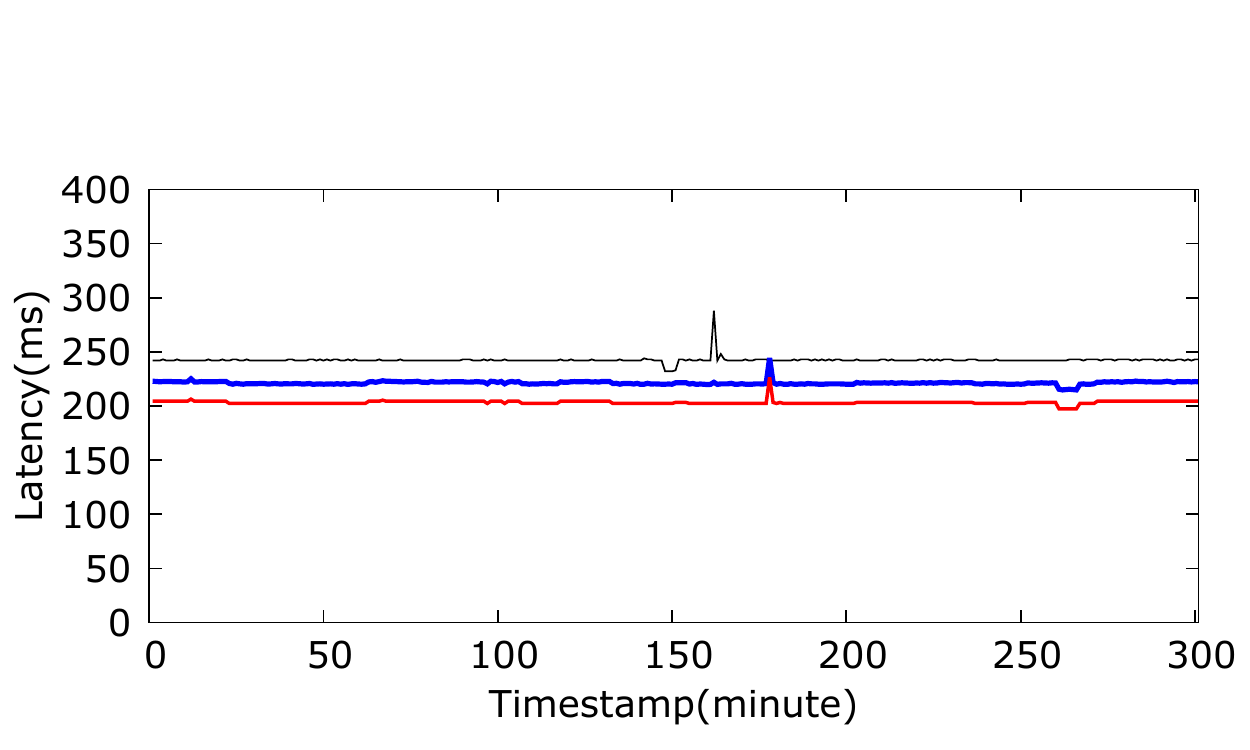}}
                              \vspace{-0.5em}

        \caption{Latency (RTT) variations of \NprojectName and the ISP-based Internet paths}
                            \vspace{-1.5em}

        \label{fig:jitter}

\end{figure*}

\smartparagraph{Latency:} 
We then benchmarked the latency and jitter of \NprojectName against that of the ISP-based internet connectivity by observing the variations in latency. We modeled data transfer from Atlanta to Sydney, Tokyo, Mumbai, and Seoul, via the cloud region of North Virginia. North Virginia was chosen based on its proximity to Atlanta, and based on our previous observation of the highest data rate. We considered two scenarios of \NprojectName in connecting our server in Atlanta to the nearest cloud region in North Virginia. First, via the ISP-based connectivity, and then via a dedicated link (such as the AWS Direct Connect) that we modeled. Figure~\ref{fig:jitter} shows the latency variations as observed during 5 hours (measured over different periods in a day), for the data flow between the 4 pairs of origin and destination.

We note that \NprojectName offers minimal jitter, with reduced variation in latency across all the destination cloud regions. On the other hand, when using ISP-based public Internet paths entirely, the jitter of the path seems to rely on the time of the experiment as well as the destination significantly. During our analysis, we observed relatively more stable cloud paths, with minimal latency and jitter. Furthermore, the public Internet paths, for most of the time, incurred more latency than \NprojectName.

We observe a higher jitter as well as a higher latency (even though still lower than using the public Internet paths entirely) when \NprojectName is used in conjunction with ISP instead of a dedicated link. This observation indicates that the connection between the origin to the nearest cloud server via the ISP contributes more to the jitter when using \NprojectName without a direct connection. Even when jitter is observed due to the variation in cloud overlay network as in Figure~\ref{fig:jitter-b}, we note that using public ISP to connect the origin and destination led to an even larger jitter. During our evaluation, the cloud path between North Virginia and Tokyo showed a higher jitter. We note that the jitter between two cloud regions depend on the regions as well as the timeframe of the experiment, and is out of control of \NprojectName.

\smartparagraph{Loss rate:}
We observed a loss rate of 1.33\% when data was transferred from Atlanta to Sydney via the public Internet paths, and 1\% when \NprojectName was used in conjunction with the ISP to connect the origin server to the cloud region. All the other regions had a 0\% loss rate in all 3 cases. In any case, the cloud paths did not contribute to the packet loss, as observed by the 0\% loss in case of the \NprojectName with direct connect in Figure~\ref{fig:jitter-a}.
\section{Related Work}
\label{sec:related_work}
In this section, we look into existing work related to \NprojectName. We can create overlay networks that span the globe on top of cloud Virtual Machines (VMs), by leveraging the cloud providers that have a global presence. Such an overlay network supported by cloud VMs as its infrastructure is known as a Cloud-Assisted Network~\cite{gharakheili2017cloud}. NetUber~\cite{kathiravelu2018moving} presents an economic and high-performance connectivity-based on Cloud-Assisted Networks. However, network softwarization approaches have not been adequately studied on a global scale for their potential benefits concerning monetary cost and performance. We must ensure that the proposed \NprojectName offers better performance and bandwidth for service workflows without additional capital and operational costs.

The prevalent demand for high data rate and low latency of the Internet applications has driven more infrastructure and service providers to distribute their resources closer to the end users~\cite{bonomi2012fog}. Latency-sensitive Internet applications~\cite{shi2016edge} such as high-frequency trading~\cite{kirilenko2011flash}, online gaming~\cite{claypool2010latency}, remote surgery~\cite{anvari2005impact}, eScience workflows~\cite{taylor2014workflows}, and the IoT~\cite{de2018application, zhang2015cloud, villari2017software} have a high demand for a quick response. These Internet applications are reaching geographically diverse locations, far from the tier-1 cities that typically host cloud data centers. With the need for low latency~\cite{ahmed2017mobile}, these Internet applications perform better when they are deployed and served from the edge~\cite{shi2016edge}, compared to cloud regions that are typically farther to the users than the edge providers. The demand for a locality-aware execution is met with an increasing number of edge providers to serve the large and geographically-distributed user base~\cite{villari2016osmotic}. Subsequently, cloud providers are also opening up more regions~\cite{zhang2010cloud} to offer better QoS to the geographically distributed users.

Software-Defined Systems (SDS)~\cite{darabseh2015sdstorage} are a set of frameworks and approaches inspired by the centralized logical control offered by the SDN. SDS brings the programmability and control of SDN to heterogeneous systems, built atop various network environments. Some SDS extend and leverage SDN as their core, while others merely follow a software-defined approach inspired by SDN. As a complete softwarization of the network systems is beyond the scope of classic SDN, more and more SDS have been built, including Software-Defined Storage~\cite{thereska2013ioflow}, Software-Defined Data Center (SDDC)~\cite{ali2013software}, Software-Defined Radio~\cite{jondral2005software}, and Software-Defined Wide Area Networks (SD-WAN)~\cite{michel2017sdn}. These SDS approaches build and manage storage, data centers, and wide area networks via a software control plane that has control over the entire system. Some SDS extend and leverage SDN as their core, while others merely follow a software-defined approach inspired by SDN. Regardless of these promising developments, existing SDS frameworks still focus on a single provider, by having a unified global view of the system. Multi-domain wide area networks, such as inter-cloud~\cite{grozev2014inter} and edge environments, require collaboration and coordination among several providers, each managing their network domain -- potentially with an SDN controller. Additional research and implementation are necessary to develop an \NprojectName for scheduling web service workflows at an Internet-scale, considering the diversity of the services that compose the user workflows.

There have also been efforts to make Internet exchanges dynamic with Software-Defined Internet Exchange (SDX)~\cite{gupta2014sdx}. Initiatives such as SDX are instrumental in establishing the Internet as a more dynamic network. However, more efforts should be taken from the perspective of the user, to schedule web service workflows at Internet scale. \NprojectName is an attempt at leveraging the application layer to schedule user workflow execution in a latency-aware manner at the Internet scale.

We should identify the potential and feasibilities for coordination and management of an entire wide area network of multi-domain environments such as the edge and multi-clouds. A global view of the whole network achieved by SDN is advantageous in managing and controlling the environments that are operated by a single entity, such as a data center or a cloud provider. While extending SDN for wide area networks has previously been proposed in frameworks such as SD-WAN~\cite{michel2017sdn}, they limit their focus to network services offered by a single provider. On the other hand, edge and inter-cloud environments consist of several network providers providing various services to multiple tenants. Consequently, a global view and control of the entire network are currently infeasible in such multi-domain environments, due to technological as well as administrative and management challenges in making coordination and collaboration across multiple providers. We need to devise an \NprojectName that offers management and coordination capabilities for multi-domain environments without sacrificing the independence and security of each domain.
	
\section{Conclusion}
\label{sec:conclusion}
The Internet has largely remained a static entity, with minimal control given to the end users on how data reaches them, or their data reaches another endpoint in the world. With the increasing number of edge providers, we ask whether an \NprojectName can be devised, to choose Internet routes with latency awareness for service workflow scheduling. While it is out of the scope of this paper to build an entire architecture replacing the Internet, we note that individual web service workflows can exploit the ever-growing edge computing nodes. We propose that the application layer can extend and exploit network softwarization for latency-aware Internet transfers at the Internet scale. Our preliminary evaluations indicate that such approaches can significantly reduce the latency and jitter of Internet web service executions.

Execution environments managed by third-party providers lack interoperability among them, thus preventing the users from exploiting resources from multiple providers. While SDN improves the configurability of the networks, heterogeneous devices, and executions on the Internet do not typically support SDN protocols. Furthermore, as a network protocol, the scope of SDN is limited to control of network data plane devices, usually belonging to a single domain such as a data center. Network softwarization must be extended to realize \NprojectName, while also confining to the limitations of Internet architecture. A complete \NprojectName architecture should be built to efficiently share resources from the diverse providers for the user web service workflow executions at an Internet scale.

\balance


\bibliography{references}

\end{document}